\documentclass[11pt]{article}

\usepackage[polish,english]{babel}
\usepackage{epsfig}

\def\L#1#2#3#4#5{#1[#2]$_{\rm #3}\;$(#4$_{\rm #5}$)}   
\def\S#1#2#3{#1[#2]$_{\rm #3}$}                        
\def\Q#1#2{#1$\pm$#2}                                  
\def\QI#1#2{{\it #1}$\pm${\it #2\/}}

\def\etal{{\em et al.\/~}}
\def\spectrum#1#2{#1$\;${\small\rm #2}\relax}

\begin{document}


%
%
\begin{center}
{\large {\bf 
Accurate transition rates for the 5p -- 5s transitions in
\spectrum{Kr}{I}  }}\\[7mm]

\selectlanguage{polish}Krzysztof
Dzier"r"ega,\selectlanguage{english}$^{\rm 1,2}$ Udo Volz,$^{\rm 3}$
Gillian Nave,$^{\rm 1,4}$ and Ulf Griesmann$^{\rm 1,4}$\\[1mm]

{\it 
$^{\rm 1}$National Institute of Standards and Technology, Gaithersburg, MD
20899--8421, U.S.A.\\
$^{\rm 2}$Instytut Fizyki im.\,M.\,Smoluchowskiego, Uniwersytet Jagiello\'nski,
30--059 Krak\'ow, Poland\\
$^{\rm 3}$Fachbereich Physik, Universit\"at Kaiserslautern,
D--67653 Kaiserslautern, Germany\\
$^{\rm 4}$Harvard--Smithsonian Center for Astrophysics, Cambridge,
MA 02138, U.S.A.\\[5mm]
}
\end{center}

%
%
\vspace{4mm}
\setlength{\parindent}{15mm}
\begin{minipage}[t]{165mm}
Branching fractions were measured for electric dipole transitions from
the 5p upper levels to the 5s levels in neutral krypton atoms. The
measurements were made with a wall-stabilized electric arc and a 2m
monochromator for the spectral lines in the visible, and with a hollow
cathode lamp and the NIST 2m--Fourier transform spectrometer for the
lines in the near infrared.  A semi-empirical calculation, based on
accurately known lifetimes for six upper levels, was used to calculate
lifetimes for which accurate measurements do not exist. This resulted
in a complete set of lifetimes for all 5p levels.  Branching fractions
and lifetimes were used to calculate transition rates for the 5p--5s
transitions. The relative uncertainties of the transition rates range
from less than 1\% for the strongest lines to about 10\% for the
weakest lines. Our data also reveal that most of the previous
measurements appear to have been affected by opacity effects in the
light sources.\\[5mm]

%
%
\noindent
32.70.Cs, 32.70.Fw
\vspace{10mm}
\end{minipage}

%
%
\section{Introduction}
\label{sec:introduction}
The transitions from the 5p levels to 5s levels in neutral Kr give
rise to the most prominent lines in the \spectrum{Kr}{I} emission
spectrum (see Fig.\,\ref{figure_levels} for a simplified energy level
diagram).  Lifetimes with sub-percent uncertainties for six of the ten
5p levels have recently been measured with beam-gas laser spectroscopy
(BGLS) by Schmoranzer and Volz \cite{Schmoranzer93} and Schmitt \etal
\cite{SPVS98}. We will show below how the six lifetimes from BGLS can
be used in an intermediate coupling calculation to predict accurately
the lifetimes of those four levels that were not accessible to
BGLS. The resulting complete set of lifetimes for the 5p levels
inspired us to make new, accurate measurements of branching fractions
for all 5p--5s transitions in \spectrum{Kr}{I} to obtain electric
dipole transition rates with uncertainties limited only by the
accuracy of the radiometric calibration.

To date, the most extensive measurements of transition rates in
\spectrum{Kr}{I} were carried out by Chang, Horiguchi and Setser
\cite{CHS80} who have measured transition rates for all 5p--5s
transitions but with an accuracy of only 30\%. Similar measurements
were made by Fonseca and Campos \cite{Fonseca79,Fonseca80} who used a
low-pressure spectral lamp as an excitation source and lifetimes
measured in an electron excitation experiment for absolute
measurements of transition rates.  A number of other experiments used
thermal plasma sources, either wall-stabilized electric arcs in the
experiments by Ernst and Schulz--Gulde \cite{Ernst78} and Brandt,
Helbig and Nick \cite{Brandt82} or a shock-tube in the experiment by
Kaschek, Ernst and B\"otticher \cite{Kaschek84}. These experiments
depended on plasma diagnostics and the transition rates have relative
uncertainties that are generally not much better than $\pm$10\% even
for strong transitions.

%
%
\section{Upper level lifetimes}
\label{sec:lifetimes}
The lifetimes of six of the ten 5p upper levels are known with
relative standard uncertainties of the order of 0.2\% or better from
recent beam-gas-laser spectroscopy (BGLS) measurements by Schmitt
\etal \cite{SPVS98}. Our primary purpose here is to find reliable
estimates for the lifetimes of the remaining four 5p levels.  We will
do this with a semi-empirical theoretical approach that is based upon
intermediate coupling theory. In addition, we have evaluated the
published experimental data to have an alternate set of
lifetimes. Although these lifetimes are less accurate than the
semi-empirical ones they provide bounds for the semi-empirical
lifetimes.

\subsection{Experimental lifetimes}
A comparison with the six reference lifetimes from BGLS \cite{SPVS98}
divides the experimental data sets from literature in two classes of
different reliability. The results from experiments employing
selective laser excitation \cite{CHS80,CGO93,WBP95} and from the only
wall-stabilized arc emission experiment \cite{Ernst78} carried out so far
generally fall (with a few explainable exceptions) into a $\pm $8\%
tolerance band around the six reference lifetimes (see
Table~\ref{DGNV99.exptau}). The pulsed-laser lifetime measurements
encountered some problems for the closely-spaced levels 2p$_8$ and
2p$_9$ due to fast collisional mixing that resulted in non-exponential
decay curves. Apart from these two levels the results agree within
$\pm$8\% with the BGLS lifetimes. A tendency towards underestimated
error bars, however, is obvious for all three experiments. The
lifetimes resulting from the arc emission experiment \cite{Ernst78} also
agree within $\pm$8\% with the BGLS results despite of an uncertainty
of $\pm$30\% the authors quote for their absolute intensity scale. The
exception here is the level 2p$_9$ for which saturation problems were
not adequately treated.

These measurements (summarized in Table~\ref{DGNV99.exptau}) are the
best measured lifetimes for the remaining four 5p levels.  The
estimated relative uncertainties are around $\pm$8\%. Other
experiments which employed pulsed electron excitation
\cite{OV67,FC78,Fonseca80} or the Hanle effect
\cite{LD73,LHM75,KRC78,GKM85} all have produced at least one result
far outside of the $\pm$8\% tolerance range and will therefore not be
considered further.

\subsection{Semi-empirical lifetimes}
The 5p levels in krypton decay exclusively (apart from some very weak
far-IR channels for the four highest 5p levels) through the
transitions of the 5p--5s array. In a typical semi-empirical
calculation for this transition array (e.g.\ Lilly \cite{Lilly76}, see
column CA in Table~\ref{DGNV99.emptau}) the wavefunctions of the
initial and final configurations $\gamma = $4p$^{\rm 5}$5p and $\gamma'
= $4p$^{\rm 5}$5s in intermediate coupling are expressed in terms of
LS-coupled wavefunctions $|\gamma LSJM\!>$:
\begin{eqnarray}
|i,JM\!>   & = & \sum_{LS} |\gamma LSJM\!>\;a(\gamma LSJ, i)\\
|f,J'M'\!> & = & \sum_{L'S'} |\gamma L'S'J'M'\!>\;a(\gamma' L' S'J',
f)
\end{eqnarray}
($i$ and $f$ denote the coupled initial and final states). The mixing
co\"efficients $a(\gamma LSJ,\cdot)$ can be determined from
experimental energies with a semi-empirical fit procedure in the
manner described by Lilly \cite{Lilly76}. Once the mixing coefficients
have been determined, the reduced dipole matrix elements
$<\!i||D||f\!>$, which are proportional to the transition
rates $A_{if}$, may be expressed in terms of the reduced
matrix elements in LS-coupling. The latter can be reduced further,
using angular momentum theory, to
\begin{equation}
<\!\gamma LSJ||D||\gamma' L' S' J'\!> = 
 \delta_{SS'} (2J+1)^{1/2}(2J'+1)^{1/2}(-1)^{L+1+S+J'}
 \left\{{SLJ \atop 1J'L'}\right\} <\!\gamma LS||D||\gamma' L'S'\!>\ .
\end{equation}
In the single electron (or Coulomb) approximation the reduced dipole
matrix element is proportional to the dipole transition moment
$\sigma_{\gamma\gamma'}$:
\begin{equation}
<\!\gamma LS||D||\gamma' L'S'\!> = (2L+1)^{1/2}\; \sigma_{\gamma\gamma'}
\end{equation}
where
\begin{equation}
\label{eq:moment}
\sigma_{\gamma\gamma'} = \sqrt{3} \int_0^{\infty} u_{\rm
5p}(r)\,er\,u_{\rm 5s}(r)\;dr 
\end{equation}
and $u_{\rm 5p}(r)$, $u_{\rm 5s}(r)$ are the radial wavefunctions of
the valence electron. In this simple semi-empirical model the relative
transition rates, and thus the lifetime ratios, depend on the
intermediate coupling coefficients $a(\gamma LSJ,\cdot)$ of the 5p and 5s
configurations, and the absolute scale is given by one single
transition moment $\sigma_{\gamma\gamma'}$ for the entire 5p--5s
transition array.

When we use the transition moment $\sigma_{\gamma\gamma'} =
3.04$\,a.u. that was obtained in the semi-empirical calculation by
Lilly \cite{Lilly76} we find that, on average, the six experimental
BGLS lifetimes can be reproduced no better than within 7\%.  The
predictions from this semi-empirical model for the remaining lifetimes
are presumably not more accurate than the recommended experimental
values are (see Table~\ref{DGNV99.exptau}). It appears unlikely that
the mixing coefficients of the 5s and 5p configurations are
responsible for the lesser accuracy of the semi-empirical lifetimes
since the reproduction of the experimental energies by the
semi-empirical intermediate coupling method is quite good
\cite{Lilly76}. The problem is the assumption of one single transition
moment $\sigma_{\gamma\gamma'}$ for the entire transition array. To
refine the semi-empirical model, we assumed $LS$-dependent transition
moments $\sigma_{\gamma\gamma'}(L,S,L',S')$ which correspond to
LS-dependent radial functions $u_{\rm 5p}(LS)$ and $u_{\rm 5s}(L'S')$
in Eq. \ref{eq:moment}. These are similar to those used, for example,
in Hartree-Fock calculations. We further assumed that spin-orbit
interaction only results in a mixture of LS-terms but not in a
modification of the radial wavefunctions.  Since a calculation of
these transition moments from first principles or from experimental
energies would not have been accurate enough for our purposes we
determined the transition moments from the six reference lifetimes
from BGLS.

In total six different non-zero transition moments are needed for the
description of the 5p-5s array. They correspond to the six allowed
transitions in $LS$-coupling (see Table~\ref{DGNV99.sigma}). In one
case, for the transition 5p~$^3$D~$\rightarrow$~5s~$^3$P, the
transition moment may be calculated directly from the lifetime of the
level 2p$_9$ since both the initial ($^3$D$_3$) and the final state
($^3$P$_2$) of the only decay channel are pure states in
$LS$-coupling. Generally, the transition moments have to be determined
by means of a nonlinear least-squares fit procedure that adjusts the
transition moments so as to get best agreement of the calculated
lifetimes with the six reference lifetimes from BGLS. The results are
summarized in Table~\ref{DGNV99.sigma}. The quoted standard
uncertainties of the semi-empirical transition moments and lifetimes
were obtained by Gaussian propagation of the uncertainties of the
reference lifetimes, the uncertainties in the energy parameters (see
\cite{Lilly76}), and the uncertainties of the contributing branching
ratios.

\begin{figure}
\begin{center}
   \epsfig{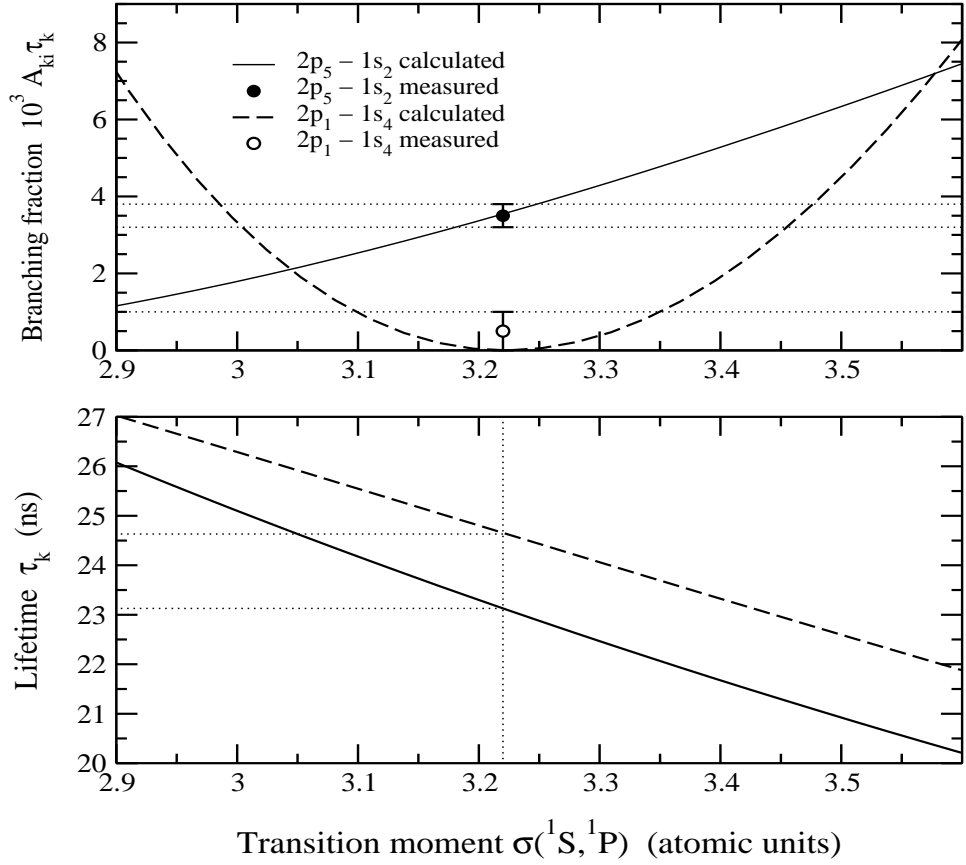}
\end{center}
\caption{Dependence of lifetimes and branching fractions of the levels
2p$_{\rm 1}$ and 2p$_{\rm 5}$ on the transition moment $\sigma$($^{\rm
1}$S,$^{\rm 1}$P). The top part of the figure shows how the measured
branching fraction determines the transition moment which, in turn,
determines the lifetime, as illustrated in the bottom part of the
figure.}
\label{figure_dep}
\end{figure}

The energy matrices and the intermediate coupling coefficients of the
5p and 5s configurations (see Table~\ref{DGNV99.emptau}) were
recalculated from the Slater- and spin-orbit parameters (including the
$\alpha L(L+1)$ correction) given by Lilly \cite{Lilly76}. The six
reference states (2p$_{3,4,6..9}$) for which the lifetimes are known
very accurately are mostly built from the four $LS$-terms $^1$P,
$^1$D, $^3$P, and $^3$D. The four corresponding transition moments
could thus be deduced with high accuracy (see
Table~\ref{DGNV99.sigma}). The transition moment of the transition
5p~$^3$S~$\rightarrow$~5s~$^3$P was determined with a somewhat greater
uncertainty from the lifetime of the 2p$_3$ state. This state is the
only one in the set of reference states that contains a relevant
contribution (16\%) from the $^3$S term. The $^1$S term only
contributes to the states 2p$_1$ and 2p$_5$ which are not included in
the set of reference states. For the determination of the transition
moment of the transition 5p~$^1$S~$\rightarrow$~5s~$^1$P we resorted
to branching fractions as additional criteria. Particularly, we used
branching fractions of the weak decay channels
2p$_1$~$\rightarrow$~1s$_4$ and 2p$_5$~$\rightarrow$~1s$_2$ which are
sensitive to the transition moment sought after.  The dependence of
the branching fractions for these transitions and the upper level
lifetimes on the transition moment is shown in figure
Fig. \ref{figure_dep}. The attainable accuracy for the transition
moment, however, is limited by the uncertainty in the branching
fractions.

As a last technical detail we note that for the four highest 5p levels
(2p$_1$ $\dots$ 2p$_4$) there are weak far-IR decay channels to states
of the 4d configuration which have to be accounted for. For this
purpose we used the theoretical transition rates calculated by Aymar
and Coulombe \cite{Aymar78}. Because of the huge discrepancies between
length- and velocity-form results for the 5p--4d transition rates we
used the greater velocity-form results with a pessimistic uncertainty
estimate of $\pm$100\% (see Table~\ref{DGNV99.emptau}).

The uncertainties of the four semi-empirical lifetimes (see
Table~\ref{DGNV99.emptau}) for the states 2p$_1$, 2p$_2$, 2p$_5$, and
2p$_{10}$ vary between 0.1\% and 3.7\% depending on the $LS$-terms
they are built from. The 2p$_2$ state allows for a very precise
lifetime calculation because it relates to the four accurately
determined transition moments only. The lifetime predictions for the
other three levels are less accurate because they include significant
contributions from the two less accurate transition moments. Our
semi-empirical predictions agree very well with the best previous
experimental values (see column BE in Table~\ref{DGNV99.exptau}) but
they are of superior accuracy and we used them for the normalization
of our transition rates.

%
%
\section{Branching fractions}
\label{sec:experiment}
\begin{figure}
\begin{center}
   \epsfig{file=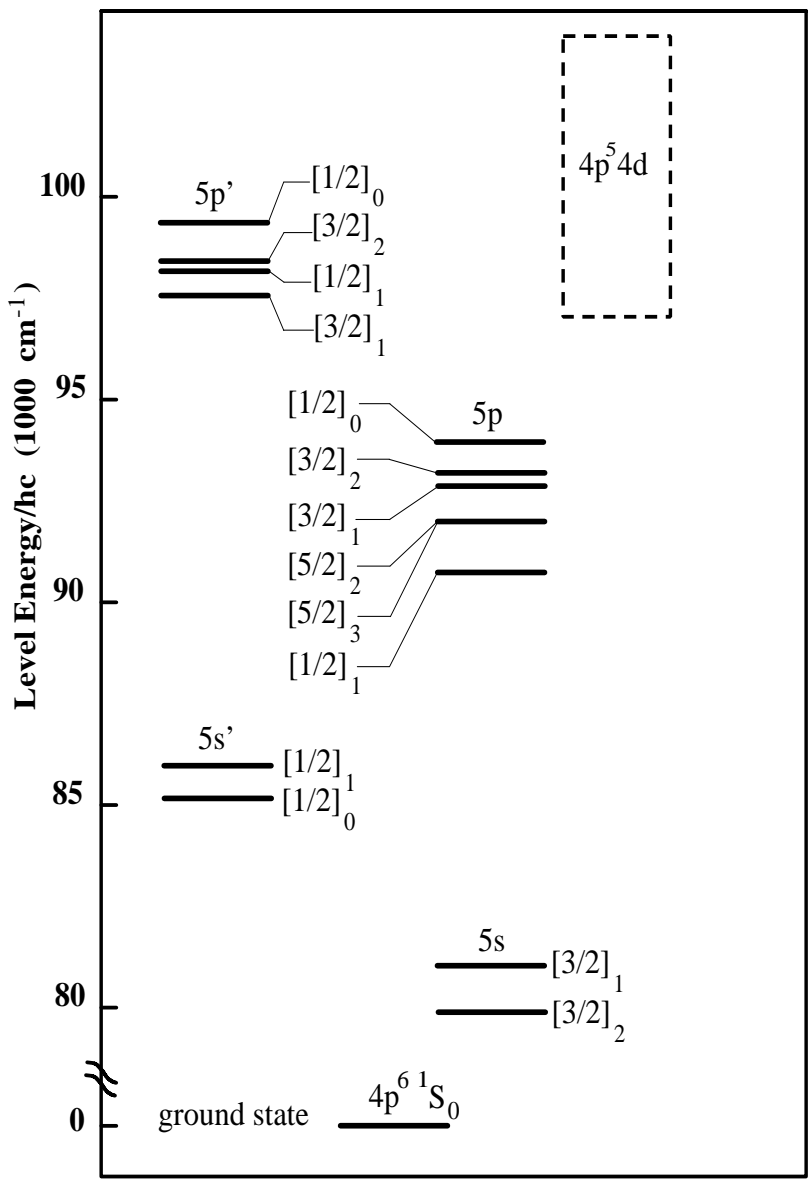, width=5in, height=7in, angle=0}
\end{center}
\caption{5s and 5p energy levels of Kr\,I. The levels are labeled in
J$_{\rm c}$K--coupling notation which is the most appropriate for the
spectra of the noble gases. The primed and unprimed levels are
distinguished by their different core angular momenta J$_{\rm c}$.
The core level of the unprimed levels is 4p$^5$\,$^2$P$_{3/2}$ whereas
it is 4p$^5$\,$^2$P$_{1/2}$ for the primed levels.  The two 5s levels
with angular momenta of 0 and 2 cannot decay into the ground state via
electric dipole transitions. Several transitions from 5p' levels to
4p$^5$\,4d levels in the infrared near 10000\,nm are outside the
wavelength range that was accessible in our experiments.}
\label{figure_levels}
\end{figure}

The measurement of branching fractions for the transition from 5p
levels presents a formidable task owing to the metastable nature of
the lower 5s and 5s$'$ levels (see Fig. \ref{figure_levels}) which may
render the lamp discharge column optically thick for transitions to
those levels.  We have measured branching fractions for 30 lines
arising from 5p levels in the wavelength range from 556.2\,nm to
1878.5\,nm in two separate experiments. The spectral lines in the
visible part of the \spectrum{Kr}{I} spectrum were measured in air
with a wall-stabilized arc discharge and a 2m -- Czerny-Turner
monochromator. The infrared portion of the spectrum was measured with
a hollow-cathode lamp and the NIST 2m -- Fourier transform
spectrometer. The comparison of the results from the two different
experiments made it easier for us to notice systematic errors due to
optically thick transitions in the light sources.  The four 5p$'$ --
4d transitions (see Fig. \ref{figure_levels}) near 10000\,nm were
outside the range of either experiment.

%
%
\subsection{Wall-stabilized arc measurements}
\label{sec:arc}
\begin{figure}
\begin{center}
   \epsfig{file=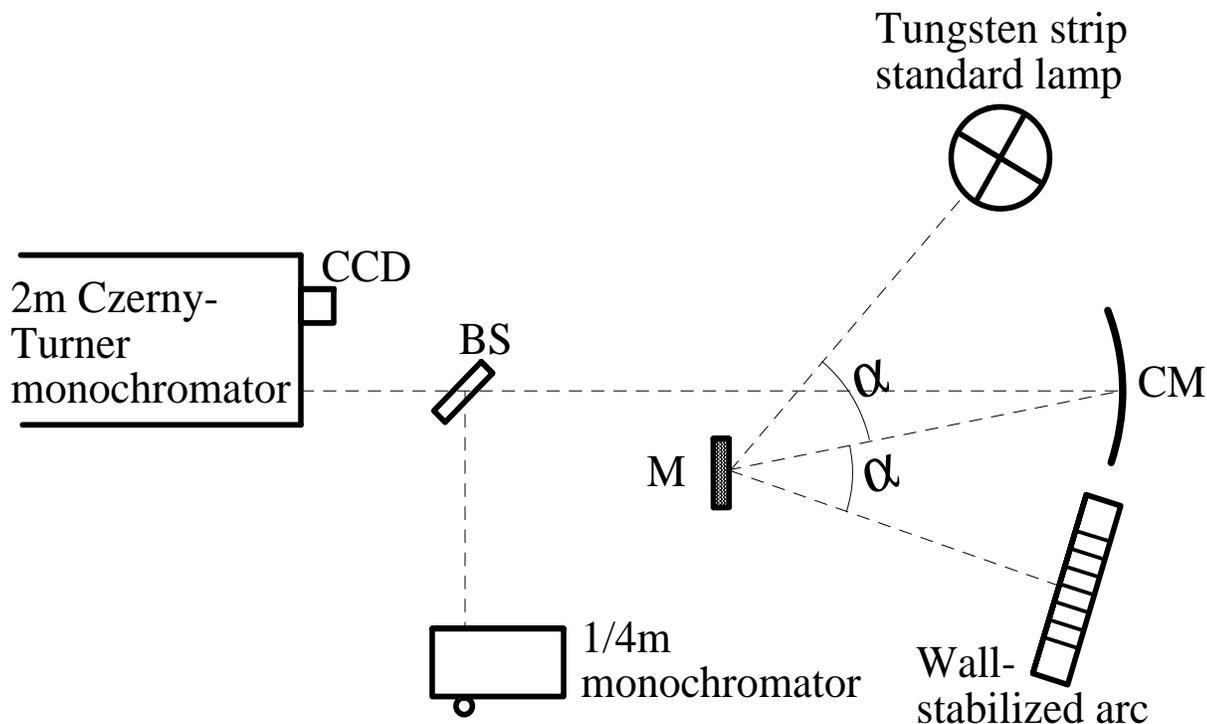, width=5in, height=7in, angle=-90}
\end{center}
\caption{Schematic of the experimental setup for the measurements with
the wall-stabilized arc. The radiation from the arc or the radiometric
standard lamp were imaged onto the entrance slit of a
2m--monochromator by a curved mirror CM. A flat mirror M, that was
mounted on a turntable allowed to alternate between both
lightsources. A small fraction of the light was imaged onto the
entrance slit of a 1/4m--monochromator with a beam splitter BS. This
monochromator remained set to the Kr\,I line at 760.2\,nm to monitor
the stability of the arc discharge. The experimental setup for the
measurements with the NIST 2m -- Fourier transform spectrometer was
identical but the imaging system was enclosed in a box that was purged
with water vapor- and carbon dioxide- free air.}
\label{figure_setup}
\end{figure}

The experimental setup for the measurements is shown schematically in
Fig. \ref{figure_setup}. In our experiment we used the wall-stabilized arc
previously described in detail by Musielok \etal
\cite{Musielok95}. The space near the electrodes was operated in argon
while the midsection of the arc channel contained helium with a small
admixture of krypton.  The fraction of krypton in helium was
maintained below 0.3\% to avoid self-absorption of krypton lines.  The
arc was operated at a current of 50\,A.  To check for optical
thickness, the krypton spectra were measured with varying amounts of
krypton in the discharge.

When the wall-stabilized arc is operated in helium, spectral lines
remain narrow and continuum emission is low because of the low
electron density in a helium arc. This facilitates more accurate line
intensity measurements because spectral lines are well isolated and
the ratio of line to continuum intensity is high. It was not necessary
to achieve LTE conditions in the arc plasma, because we were only
interested in the measurement of branching ratios of spectral
lines. The measurements were performed in a side-on configuration to
avoid interloping argon lines and argon plasma continuum radiation
that are emitted at the ends of the arc.  As indicated in Fig.\,
\ref{figure_setup}, either the wall--stabilized arc or a tungsten
strip standard lamp were imaged onto the entrance slit of a 2\,m
Czerny--Turner monochromator by a concave mirror with a magnification
factor of approximately 1.3.  A beam splitter was placed in the beam
path to reflect a fraction of the light into the 0.25\,m monochromator
that was used to monitor the discharge stability. This monochromator
was set to the 760.2\,nm line of \spectrum{Kr}{I}. The total intensity
of this line was measured with a photomultiplier tube and a chart
recorder and showed less than 1\% fluctuation during our measurements.
The krypton spectra were recorded with a CCD camera that was mounted
at the exit plane of the monochromator. The measured spectral line
profiles were first corrected for the spectral response of the
experimental system, as determined with the standard lamp, and the
residual continuum was subtracted. The lines were then integrated by
fitting a spline function to the data using a program package
published by Renka\cite{Renka87} which yields the integral of the
spectral line without requiring that the apparatus function be known
analytically.

%
%
\subsection{Hollow cathode lamp measurements}
\label{sec:hollow}
The experiment described in the previous section was unsuited for
measurements in the infrared because it was set up in air. A second
experiment in a purged environment was therefore carried out to
measure the intensity of lines in the infrared. This used a
high-resolution Fourier transform spectrometer to observe spectra of a
hollow cathode lamp.

The high-current hollow cathode lamp we used was developed by Danzmann
\etal \cite{Danzmann88}. For our measurements it was equipped with a
cathode made of oxygen-free copper which is easy to operate and has no
lines that blend with the krypton lines of interest.  The hollow
cathode lamp was operated with between 130\,Pa and 250\,Pa of argon or
neon as a carrier gas for the discharge with an admixture of between
0.5\,Pa and 10\,Pa.  The discharge current was varied between 100\,mA
and 500\,mA. The experimental setup was similar to the one used with
the wall-stabilized arc.  The entire imaging system was enclosed in a
purge box that was continuously purged with water vapor and carbon
dioxide free air to suppress absorption by these gases in the near
infrared.

Many lines were strongly self-absorbed when pure krypton was used as a
carrier gas in the hollow cathode lamp discharge.  This problem was
partly overcome when the partial pressure of krypton in the discharge
was reduced by using a neon-krypton mixture in the hollow cathode
lamp. We also found that the spectra obtained with high currents where
the copper density in the discharge is high show self absorption only
in the very strongest lines.  We assume that the metastable 5s states
were depopulated by charge-transfer collisions with copper atoms in
the hollow cathode lamp discharge.

The NIST 2m -- Fourier transform spectrometer (described in Nave \etal
\cite{Nave97}) was used to measure the spectra of the hollow cathode
lamp and the standard lamp.  A resolution of around 0.01\,cm$^{-1}$
was used for the measurements of the krypton spectra.  For the near IR
region, a liquid nitrogen cooled indium-antimonide detector was used
whereas silicon photodiodes were used to record spectra below
1000\,nm. To improve the signal-to-noise ratio, colored glass filters
were employed to restrict the bandpass of the spectrometer to the
wavenumber range of interest.  The spectral sensitivity of the optical
and detection systems was calibrated with the standard lamp before and
after measurements of the spectra of the light from the hollow cathode
discharge.  Some residual self-absorption was evident for the
strongest lines even at low krypton partial pressures. For those lines
we relied on the results from the experiment with the wall-stabilized
arc, where these lines remained optically thin.

%
%
\subsection{Data analysis and uncertainties}
\label{sec:uncertainty}
It is common that the the uncertainty of experimental transition
rates is limited by the uncertainty of the measurement of the
upper level lifetimes and not by the uncertainty of the branching
fraction measurement. In our case, the situation is reversed. The
uncertainty of the branching fraction measurement is limited by the
uncertainty of the radiometric calibration which is around 2\%. The
uncertainties of the upper level lifetimes are generally much
lower. In this section we will describe in detail how the
uncertainties for the branching fractions were calculated.

The transition rate $A_{ki}$ of a transition from a particular
upper level $k$ to lower level $i$ can be calculated from a
measurement of the upper level lifetime $\tau_k$ and a measurement of
the branching fraction $F_{ki}$ -- the fraction that the transition to
$i$ contributes to the total decay rate:
\begin{equation}\label{Avaleqn}
A_{ki} = \frac{1}{\tau_k} F_{ki} \mathrm{,\;\;\; where \;\;\; } 
F_{ki} = \frac{A_{ki}}{\sum_{j}A_{ki}}.
\end{equation}
The branching fractions can in turn be calculated from the relative
intensities $I_{ki}$ of the lines (in photons/s) by
\begin{equation}\label{BFeqn}
F_{ki} = \frac{I_{ki}}{\sum_{j} I_{ki}}
\end{equation}
where the sum is over all the lower levels to which the upper level
can decay.

Several independent measurements of the \spectrum{Kr}{I} spectrum were
made with different operating conditions for the hollow cathode lamp.
The relative intensity $\hat{I}^\alpha_{ki}$ of each spectral line in
each measured spectrum $\alpha$ was calculated from the observed
intensity $I^{\alpha}_{ki}$ and the relative efficiency of the
spectrometer $\epsilon(\sigma)$ at the wavenumber $\sigma$ of the
spectral line by:
\begin{equation}
\hat{I}^{\alpha}_{ki} = \frac{I^\alpha_{ki}(\sigma)}{\epsilon(\sigma)}
\end{equation}
where the relative efficiency of the optical system was assumed to be
constant over the width of the spectral line. These intensities were
then divided by a normalizing factor $\hat{I}^\alpha_{\rm norm}$ to
put all the intensities in all spectra on the same relative intensity
scale. This normalizing factor was usually chosen such that the
intensity of one strong line common to all spectra was 1, hence making
the intensities relative with respect to that strong line. This
approach was found to be more reliable than using a weighted mean of
the intensities, as lines in some of the spectra may be affected by
self-absorption, or be too weak to be measured. The weighted mean
relative intensity of the line $\bar{I}_{ki}$ was then found using:
\begin{equation}\label{BReqn}
\bar{I}_{ki} = \frac{1}{\sum_{\alpha} w^{\alpha}_{ki} }\sum_{\alpha}
\frac{w^{\alpha}_{ki}\hat{I}^{\alpha}_{ki}}{\hat{I}^{\alpha}_{\rm norm}}
\end{equation}
where w$^{\alpha}_{ki}$ is a weighting factor. The weighting factor
chosen for the hollow cathode measurements was the signal-to-noise
ratio of the line.  The small uncertainty of the normalization line
intensity is due to its high signal-to-noise ratio.  These branching
ratios are converted to branching fractions and transition rates using
equations \ref{Avaleqn} and \ref{BFeqn}.

Absolute transition rates are then determined from the mean values of
between 5 and 9 independent measurements of the relative intensities
and experimental lifetime data. They are presented in table
\ref{table_aik}, along with the lifetimes of the upper levels used to
determine the transition rates. The uncertainties given in the table
result from the estimated standard deviation of the branching
fractions and the uncertainty of the lifetime data. The estimated
standard deviation of the branching fractions depends on the
uncertainty in the weighted mean relative intensity, which in turn
depends on the individual measurements of the intensity through
equation \ref{BReqn}, and the uncertainty in the radiometric
calibration of the spectrometer.

The estimated uncertainty in the individual measurements of the
intensity was taken as the intensity divided by the signal-to-noise
ratio: $\hat{I}^{\alpha}_{ki}/w^\alpha_{ki}$.  When photon noise is
the dominant source of uncertainty, the square of the signal-to-noise
ratio must be used as the weighting factor in equation \ref{BReqn}.
We chose to weight the individual intensity measurements with the
signal-to-noise ratio to account for a significant systematic
component in the uncertainty which may result from self-absorption or
line blends.

The statistical component in the uncertainty of the weighted
mean relative intensity $u_{stat}(\hat{I}_{ki}$) can then be derived
by applying the law of propagation of uncertainty to equation
\ref{BReqn}:
\begin{displaymath}
u_{\rm stat}(\bar{I}_{ki}) = \sqrt{
\sum_{\alpha} \left( \frac{\hat{I}^{\alpha}_{ki}}{\hat{I}^{\alpha}_{\rm norm}} 
\cdot \frac{1}{w^{\alpha}_{ki}} \right)^2 }
\end{displaymath}
where the sum is again over all the observations of the lines.  This
must be added in quadrature to the uncertainty in the radiometric
calibration of the spectrometer, which was estimated at 3.3\% for one
standard deviation. This estimate includes the uncertainty in the
supplied calibration of the standard lamp (1.5\% for one standard
deviation) and a contribution of 3\% for the measurement of the
standard lamp spectrum. The total uncertainty in the measurement of
the weighted mean relative intensities is thus:
\begin{equation}
u(\bar{I}_{ki}) = \sqrt{u_{\rm stat}^2(\bar{I}_{ki}) + (0.033 \bar{I}_{ki})^2}
\end{equation}
The uncertainty in the measurement of the branching fractions
$u(F_{ki}$) is derived by applying the law of propagation of
uncertainty to equation \ref{BFeqn} to give:
\begin{equation}
u(F_{ki}) = \sqrt{ 
\frac{u^2(\bar{I}_{ki})}{(\sum_i \bar{I}_{ki})^2} +
\frac{\bar{I}^2_{ki}} {(\sum_i \bar{I}_{ki})^4}\sum_ju^2(\bar{I}_{ki})}
\end{equation}
This is combined in quadrature with the uncertainty in the lifetime
$u(\tau_k$) to give the uncertainty in the transition rate
$u(A_{ki}$): 
\begin{equation}
u(A_{ki}) = \sqrt{
\frac{1}{\tau^2_k} u^2(F_{ki}) + 
\frac{F^2_{ki}}{\tau^4_k} u^2(\tau_k)}
\end{equation}
For the wall-stabilized arc measurements, the transition rates
and their uncertainties were calculated similarly.

%
%
\section{Discussion of results}
\label{sec:results}
Our new transition rates for 5p -- 5s transitions in \spectrum{Kr}{I}
are listed in Table \ref{table_aik}. Also listed in Table
\ref{table_aik} are our experimental branching fractions and the
lifetimes that were used to calculate the transition rates.

\begin{figure}
\begin{center}
   \epsfig{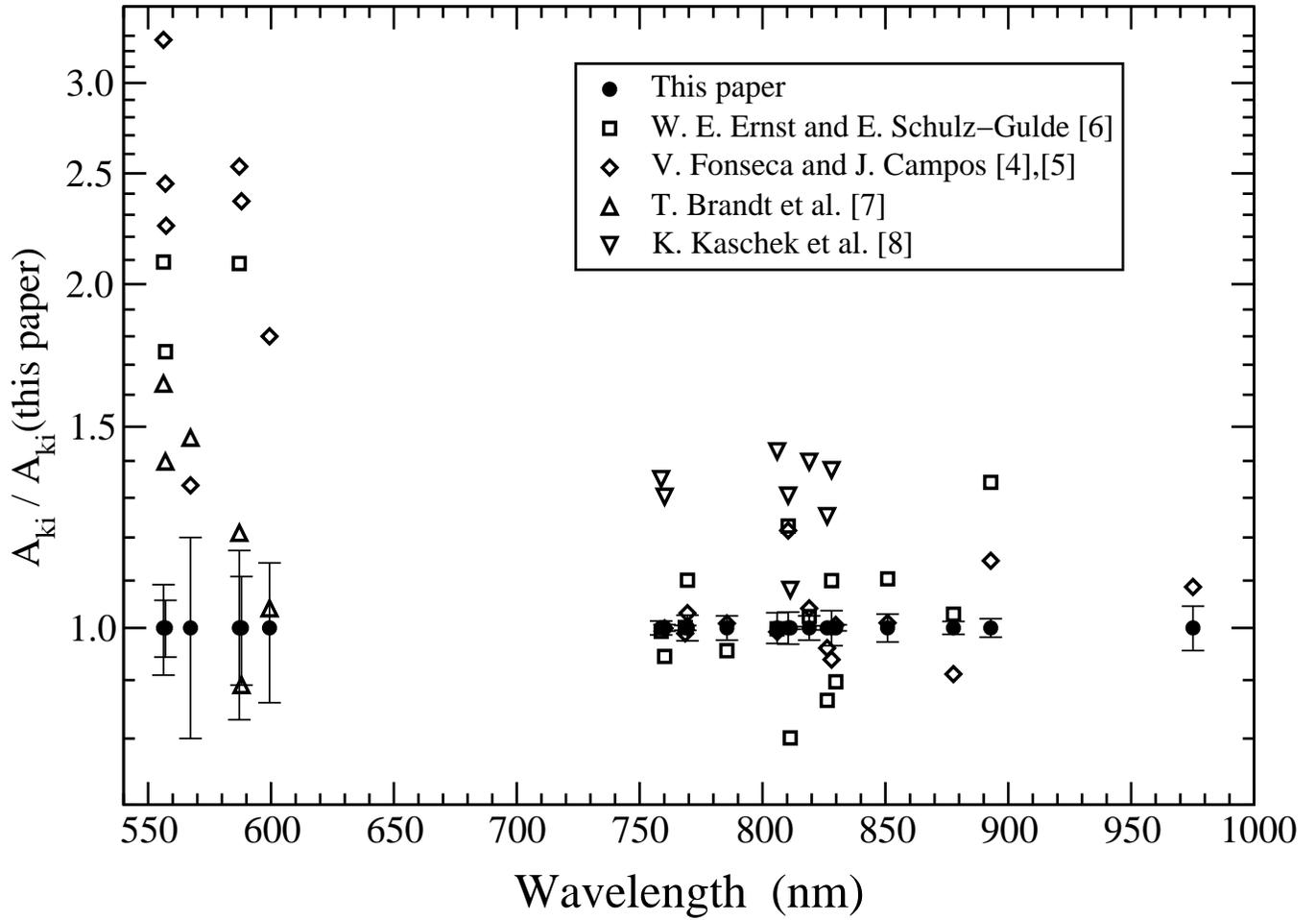}
\end{center}
\caption{Ratio of transition rates from several earlier
experiments and our results.}
\label{figure_exp}
\end{figure}

\begin{figure}
\begin{center}
   \epsfig{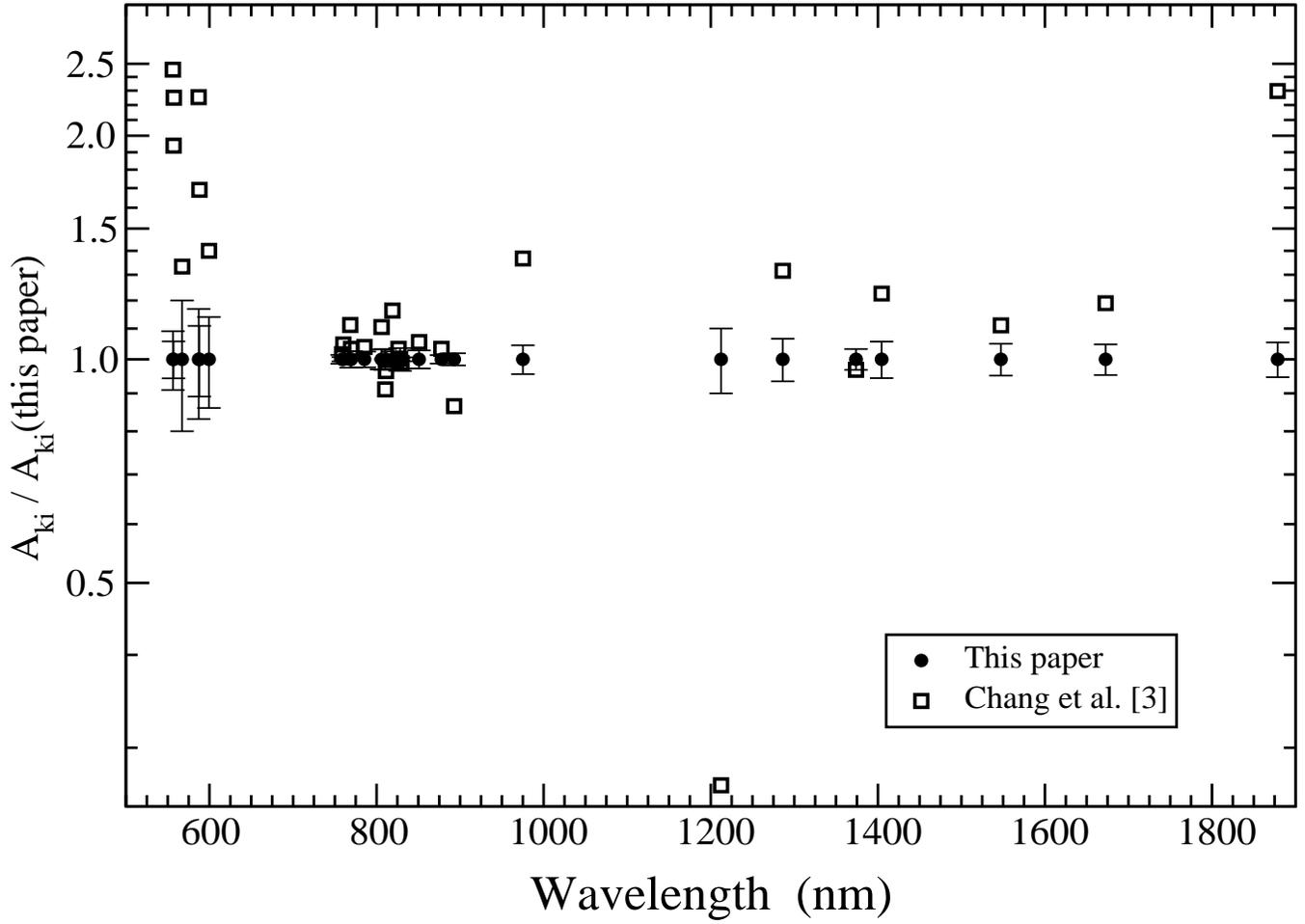}
\end{center}
\caption{A comparison of our transition rates with the
complete set of 5p--5s transition rates measured by Chang,
Horiguchi and Setser \cite{CHS80}.}
\label{figure_all}
\end{figure}

In Table \ref{table_comp} and Fig.\,\ref{figure_exp} we compare our
transition rates with several experimental results. The only
other measurement that includes the lines in the IR is that of Chang,
Horiguchi and Setser \cite{CHS80} and a comparison of those results
with our data is shown in Fig.\,\ref{figure_all}.  The results by
Fonseca and Campos \cite{Fonseca79,Fonseca80}, presented in
Fig. \ref{figure_exp}, were recalculated using the same lifetime data as
in our work.  For the set of strong lines around 800\,nm our
transition rates are in good agreement with most of the
results obtained by Fonseca and Campos \cite{Fonseca79,Fonseca80} and
differ from those obtained by Ernst \etal \cite{Ernst78} by about
10\%. The results of Kaschek \etal \cite{Kaschek84} differ from ours
by a constant scaling factor of 1.3, on average.  Only in the case of
the line at 810.4\,nm, all other experimental data exceed our result
by 20\%--30\%.

The most striking difference between previous measurements and our
results is that our transition rates for the set of weak lines
near 600\,nm are much lower than all previous measurements with the
exception of the experiment by Brandt, Helbig and Nick
\cite{Brandt82}. This strongly suggests that many of the earlier
experiments had problems with self-absorption of the strong lines
around 800\,nm which would make the weak lines in a set of transitions
from a particular upper level appear stronger.

\begin{figure}
\begin{center}
   \epsfig{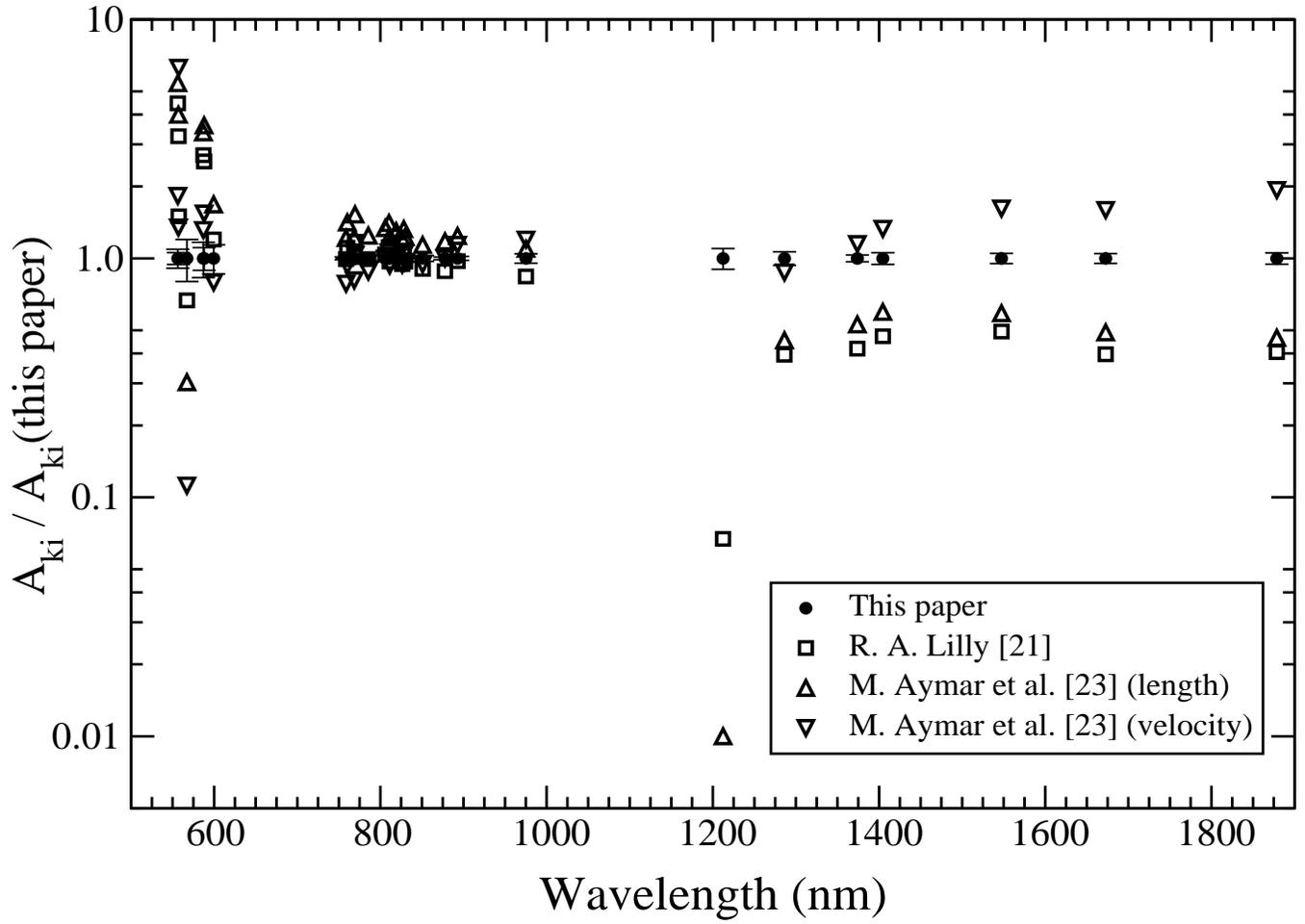}
\end{center}
\caption{A comparison of our transition rates with
theoretical transition rates.}
\label{figure_theo}
\end{figure}

It is also interesting to compare our results with the theoretical
calculations because all calculations were intermediate-coupling
calculations in the Coulomb approximations whereas our semi-empirical
lifetimes were obtained with a modified intermediate-coupling scheme.
Fig.\,\ref{figure_theo} compares our results to the most recent
calculations.  The earlier calculations by Murphy \cite{Murphy68} are
not included because they were superseded by those of Lilly
\cite{Lilly76}.  For the strongest lines near 800\,nm the best
agreement, within 10\% on average, was found between our data and
calculations made by Aymar and Coulombe \cite{Aymar78} with a velocity
dipole operator, while there is a constant disagreement (a factor of
1/3) when they used a length dipole operator.  A similar discrepancy
was found with calculations made by Lilly \cite{Lilly76}. We note that
the discrepancies for the weak lines near 600\,nm and in the IR are
considerable but there appear to be no conspicuous systematic trends
as we found in the experimental data.

%
%
\vspace*{10mm}
\begin{center}
{\large {\bf Acknowledgments}}
\end{center}
K. \selectlanguage{polish}Dzier"r"ega\selectlanguage{english}
gratefully acknowledges financial support from the Maria
Sk{\l}odowska-Curie Foundation through grant number
MEN--NIST--96--260. We are grateful to W. L. Wiese, NIST,
K.\,Musio{\l}, Jagiellonian University, Krak\'{o}w, and
H. Schmoranzer, University of Kaiserslautern for helpful discussions
and continued support.  G. Nave and U. Griesmann were supported by
NIST contracts number 43SBNB867005 and 43SBNB960002 to Harvard College
Observatory.

%
%

%
%
\newpage
\begin{table}
\begin{center}
\caption{Experimental lifetimes of the KrI 5p states. (Uncertainties
are given in parentheses.)}
\label{DGNV99.exptau}
\vspace{6pt}
\begin{tabular}{llcccccc}
\hline
\hline
\multicolumn{2}{c}{Level} & 
\multicolumn{5}{c}{Experimental lifetimes (ns)} \\
&& WSA$^1$ \cite{Ernst78}& PL$^2$ \cite{CHS80}& PL$^2$ \cite{CGO93,WBP95}& 
BGLS$^3$ \cite{Schmoranzer93} &BGLS$^3$ \cite{SPVS98} & BE$^4$\\
\hline
2p$_1$ & 5p$'$[1/2]$_0$ & 24.6(1.5) & ---       & ---       & ---       & ---       & 24.6(2.0) \\
2p$_2$ & 5p$'$[3/2]$_2$ & 32.0(1.9) & 26.9(0.3) & ---       & ---       & ---       & 29.5(2.4) \\
2p$_3$ & 5p$'$[1/2]$_1$ & 27.2(1.6) & 26.8(1.7) & ---       & ---       & 28.075(30)&           \\
2p$_4$ & 5p$'$[3/2]$_1$ & 27.9(1.7) & 27.2(1.6) & ---       & ---       & 29.402(42)&           \\
2p$_5$ & 5p [1/2]$_0$ & 23.4(1.4)   & ---       & 23.5(1.0) \cite{WBP95}& ---       & ---       & 23.5(1.9) \\
2p$_6$ & 5p [3/2]$_2$ & 28.6(1.7) & 25.4(0.8)   & 26.4(0.5) \cite{WBP95}& 27.35(6)  & 27.345(16)&           \\
2p$_7$ & 5p [3/2]$_1$ & 32.3(1.9) & 29.7(1.0)   & ---       & 29.51(6)  & 29.619(17)&            \\
2p$_8$ & 5p [5/2]$_2$ & 29.6(4.6) & 26.5(2.0)   & 32.5(0.8) \cite{CGO93}& 32.10(9)  & 32.041(47)&             \\
2p$_9$ & 5p [5/2]$_3$ & 34.6($^{+2.2}_{-5.8})$  & 28.7(2.0) & 30.2(1.4) \cite{CGO93}& 27.73(7)  & 27.694(18)& \\
2p$_{10}$ & 5p [1/2]$_1$ & ---    & 40.9(1.7)   & ---       & ---       & ---       & 40.9(3.3) \\
\hline
\hline
\end{tabular}
\end{center}
$^1$Wall-stabilized arc emission. Quoted uncertainties do not include
the uncertainty of the absolute intensity scale for the transition
rates of $\pm$30\%. \\ 
$^2$Lifetimes from pulsed laser excitation.\\ 
$^3$Lifetimes from beam-gas-laser spectroscopy.\\ 
$^4$Best estimate of non-BGLS experimental lifetimes.\\
\end{table}

\begin{table}
\begin{center}
\caption{Semi-empirical lifetimes and decomposition in $LS$-terms for the KrI 5p states.}
\label{DGNV99.emptau}
\vspace{6pt}
\begin{tabular}{llccccccccc}
\hline
\hline
\multicolumn{2}{c}{Level} & 
\multicolumn{6}{c}{$LS$ decomposition$^1$ (\%)} & Far-IR & 
\multicolumn{2}{c}{Lifetimes (ns)} \\
&& $^1$S & $^1$P & $^1$D & $^3$S & $^3$P & $^3$D &
branches$^2$ (\%) & CA$^3$ \cite{Lilly76} & this work$^4$ \\
\hline
2p$_1$ & 5p$'$[1/2]$_0$ & 43  & --- & --- & --- & 57  & --- & 0.246 & 24.37 & 24.58 $\pm$ 0.85 \\
2p$_2$ & 5p$'$[3/2]$_2$ & --- & --- & 34  & --- & 21  & 44  & 0.034 & 29.10 & 28.59 $\pm$ 0.07 \\ 
2p$_3$ & 5p$'$[1/2]$_1$ & --- & 19  & --- & 16  & 61  & 3   & 0.106 & 26.35 & 28.08 $\pm$ 0.07 \\ 
2p$_4$ & 5p$'$[3/2]$_1$ & --- & 24  & --- & 1   & 1   & 74  & 0.0045& 30.65 & 29.34 $\pm$ 0.04 \\ 
2p$_5$ & 5p [1/2]$_0$   & 57  & --- & --- & --- & 43  & --- & ---   & 23.54 & 23.12 $\pm$ 0.85 \\
2p$_6$ & 5p [3/2]$_2$   & --- & --- & 22  & --- & 75  & 4   & ---   & 25.43 & 27.34 $\pm$ 0.02 \\
2p$_7$ & 5p [3/2]$_1$   & --- & 53  & --- & --- & 24  & 23  & ---   & 30.09 & 29.62 $\pm$ 0.02 \\
2p$_8$ & 5p [5/2]$_2$   & --- & --- & 44  & --- & 4   & 52  & ---   & 34.03 & 32.05 $\pm$ 0.06 \\ 
2p$_9$ & 5p [5/2]$_3$   & --- & --- & --- & --- & --- & 100 & ---   & 28.77 & 27.70 $\pm$ 0.02 \\
2p$_{10}$ & 5p [1/2]$_1$& --- & 4   & --- & 83 & 14   & --- & ---   & 40.30 & 38.13 $\pm$ 1.07 \\ 
\hline
\hline
\end{tabular}
\end{center}
$^1$Recalculated from the parameters given by Lilly \cite{Lilly76}.\\
$^2$Total far-IR (5p-4d) decay rate over total decay rate.  Estimated
uncertainty $\pm$100\%.\\ 
$^3$Intermediate coupling with transition
moment from Coulomb approximation. \\ 
$^4$Intermediate coupling with $LS$-dependent transition moments from
Table~\ref{DGNV99.sigma}. Note the excellent agreement of these
semi-empirical lifetimes with the experimental BGLS lifetimes in Table
\ref{DGNV99.exptau}.
\end{table}

\begin{table}
\begin{center}
\caption{$LS$-dependent transition moments for the KrI 5p-5s
transition array.}
\label{DGNV99.sigma}
\begin{tabular}{lcll}
\hline
\hline
\multicolumn{3}{c}{Transition} & Transition moment $\sigma$ (a.u.) \\
\hline
5p~$^1$S & $\longrightarrow$ & 5s~$^1$P & 3.22  $\pm$ 0.18  \\
5p~$^1$P & $\longrightarrow$ & 5s~$^1$P & 3.146 $\pm$ 0.003 \\
5p~$^1$D & $\longrightarrow$ & 5s~$^1$P & 3.205 $\pm$ 0.014 \\
5p~$^3$S & $\longrightarrow$ & 5s~$^3$P & 3.173 $\pm$ 0.057 \\
5p~$^3$P & $\longrightarrow$ & 5s~$^3$P & 2.853 $\pm$ 0.073 \\
5p~$^3$D & $\longrightarrow$ & 5s~$^3$P & 3.085 $\pm$ 0.001 \\
\hline
\hline
\end{tabular}
\end{center}
\end{table}

\begin{table}
\begin{center}
\caption{ Branching fractions $F_{ki}$ and absolute transition
rates $A_{ki}$ for all 30 5p -- 5s transitions in neutral
Kr.}
\label{table_aik}
\begin{tabular}{lllllll}
\hline
\hline
Upper level & Lower level  & $\lambda$\,(nm) & 
$\tau$\,(ns)$^1$ & $F_{ki}$(WSA)$^2$ & $F_{ki}$(HCL)$^3$
&$A_{ki}$(10$^6$\,s$^{-1})$ \\
\hline
\L{5p'}{1/2}{0}{2p}{1} & \L{5s}{3/2}{1}{1s}{4}  & 557.31 & \QI{24.58}{0.85} & $<$10$^{-3}$       &                  &  $<$0.04         \\
                       & \L{5s'}{1/2}{1}{1s}{2} & 768.52 &                  & $>$0.999           &                  & \Q{40.64}{0.2}   \\[2mm]
														                       
\L{5p'}{3/2}{2}{2p}{2} & \L{5s}{3/2}{2}{1s}{5}  & 556.22 & \QI{28.59}{0.07} & \Q{0.003}{0.004}   &                  & \Q{0.11}{0.01}   \\
                       & \L{5s}{3/2}{1}{1s}{4}  & 587.09 &                  & \Q{0.020}{0.004}   &                  & \Q{0.71}{0.14}   \\
                       & \L{5s'}{1/2}{1}{1s}{2} & 826.32 &                  & \Q{0.977}{0.005}   &                  & \Q{34.16}{0.19}  \\[2mm]
														                       
\L{5p'}{1/2}{1}{2p}{3} & \L{5s}{3/2}{2}{1s}{5}  & 557.03 & \Q{28.075}{0.03} & \Q{0.028}{0.0016}  &                  & \Q{0.98}{0.056}  \\
                       & \L{5s}{3/2}{1}{1s}{4}  & 587.99 &                  & \Q{0.002}{0.0002}  &                  & \Q{0.055}{0.006} \\
                       & \L{5s'}{1/2}{0}{1s}{3} & 785.48 &                  & \Q{0.573}{0.008 }  &                  & \Q{20.41}{0.5}   \\
                       & \L{5s'}{1/2}{1}{1s}{2} & 828.11 &                  & \Q{0.398}{0.014 }  &                  & \Q{14.18}{0.5}   \\[2mm]
														                       
\L{5p'}{3/2}{1}{2p}{4} & \L{5s}{3/2}{2}{1s}{5}  & 567.25 & \Q{29.402}{0.042}& \Q{0.00044}{0.0001}&                  & \Q{0.015}{0.003} \\
                       & \L{5s}{3/2}{1}{1s}{4}  & 599.39 &                  & \Q{0.0015}{0.0002 }&                  & \Q{0.050}{0.007} \\
                       & \L{5s'}{1/2}{0}{1s}{3} & 805.95 &                  & \Q{0.465}{0.014}   &                  & \Q{15.83}{0.49}  \\
                       & \L{5s'}{1/2}{1}{1s}{2} & 850.89 &                  & \Q{0.533}{0.015}   &                  & \Q{18.11}{0.51}  \\[2mm]
														                       
\L{5p}{1/2}{0}{2p}{5}  & \L{5s}{3/2}{1}{1s}{4}  & 758.74 & \QI{23.12}{0.85} &                    &\Q{0.9965}{0.0033}& \Q{43.10}{0.6}   \\
                       & \L{5s'}{1/2}{1}{1s}{2} & 1212.35&                  &                    &\Q{0.0035}{0.0003}& \Q{0.15}{0.015}  \\[2mm]
														                       
\L{5p}{3/2}{2}{2p}{6}  & \L{5s}{3/2}{2}{1s}{5}  & 760.15 & \Q{27.345}{0.016}& \Q{0.743}{0.007}   &\Q{0.751}{0.008}  & \Q{27.32}{0.18}  \\
                       & \L{5s}{3/2}{1}{1s}{4}  & 819.01 &                  & \Q{0.248}{0.009}   &\Q{0.241}{0.008}  & \Q{8.94}{0.22}   \\
                       & \L{5s'}{1/2}{1}{1s}{2} & 1373.89&                  &                    &\Q{0.0083}{0.0003}& \Q{0.31}{0.01}   \\[2mm]
														                       
\L{5p}{3/2}{1}{2p}{7}  & \L{5s}{3/2}{2}{1s}{5}  & 769.45 & \Q{29.619}{0.017}& \Q{0.127}{0.004 }  &\Q{0.127 }{0.004 }& \Q{4.27}{0.11}   \\
                       & \L{5s}{3/2}{1}{1s}{4}  & 829.81 &                  & \Q{0.868}{0.005 }  &                  & \Q{29.31}{0.18}  \\
                       & \L{5s'}{1/2}{0}{1s}{3} & 1286.19&                  &                    &\Q{0.0031}{0.0002}& \Q{0.076}{0.005} \\
                       & \L{5s]}{1/2}{1}{1s}{2} & 1404.57&                  &                    &\Q{0.0023}{0.0001}& \Q{0.106}{0.006} \\[2mm]
														                       
\L{5p}{5/2}{2}{2p}{8}  & \L{5s}{3/2}{2}{1s}{5}  & 810.44 & \Q{32.041}{0.047}&\Q{0.288 }{0.015}   &\Q{0.287}{0.009}  & \Q{8.96}{0.29}   \\
                       & \L{5s}{3/2}{1}{1s}{4}  & 877.68 &                  &\Q{0.709 }{0.01 }   &\Q{0.710}{0.009}  & \Q{22.17}{0.29}  \\
                       & \L{5s'}{1/2}{1}{1s}{2} & 1547.40&                  &                    &\Q{0.0026}{0.0001}& \Q{0.081}{0.004} \\[2mm]
														                       
\L{5p}{3/2}{3}{2p}{9}  & \L{5s}{3/2}{2}{1s}{5}  & 811.29 & \Q{27.694}{0.018}&  1.00              &  1.00            & \Q{36.1}{0.09}   \\[2mm]
														                       
\L{5p}{1/2}{1}{2p}{10} & \L{5s}{3/2}{2}{1s}{5}  & 892.87 & \QI{38.13}{1.07} &\Q{0.873 }{0.004 }  &                  & \Q{22.89}{0.65}  \\
                       & \L{5s}{3/2}{1}{1s}{4}  & 975.18 &                  &\Q{0.120 }{0.004 }  &\Q{0.120}{0.004}  & \Q{3.13}{0.14}   \\
                       & \L{5s'}{1/2}{0}{1s}{3} & 1672.65&                  &                    &\Q{0.0048}{0.0002}& \Q{0.126}{0.006} \\
                       & \L{5s'}{1/2}{1}{1s}{2} & 1878.55&                  &                    &\Q{0.0028}{0.0001}& \Q{0.074}{0.003} \\
\hline
\hline
\end{tabular}
\end{center}
$^1$Lifetimes of the upper levels are experimental lifetimes from Schmitt \etal \cite{SPVS98}
(see also Table \ref{DGNV99.exptau}) and semi-empirical lifetimes from
Table \ref{DGNV99.emptau}. Semi-empirical lifetimes are printed in italics.\\
$^2$Wall-stabilized arc measurement.\\
$^3$Hollow-cathode lamp measurement. 
\end{table}

\begin{table}
\begin{center}
\caption{Compilation of experimental and theoretical transition rates for 5p -- 5s 
transitions in \spectrum{Kr}{I}.}
\label{table_comp}
\begin{tabular}{llllllllllll}
\hline
\hline
Upper level & Lower level  & $\lambda$\,(nm) &
$A_{ki}$(10$^6$\,s$^{-1})$ $^{\rm 1}$ &
\cite{Ernst78} & \cite{Fonseca79,Fonseca80} & \cite{Brandt82} & \cite{Kaschek84} & 
\cite{CHS80} & \cite{Lilly76} & \cite{Aymar78}$^{\rm 2}$ & \cite{Aymar78}$^{\rm 3}$ \\
\hline
\S{5p'}{1/2}{0} & \S{5s}{3/2}{1}  & 557.31 &  $<$0.04 &         &   0.09  &         &        &   0.09  &   0.06  &  0.437  &  0.251     \\
                & \S{5s'}{1/2}{1} & 768.52 & 40.64(20)&   40.7  &   40.2  &         &        &   45.2  &   41.19 & 49.6    &  32.9      \\[2mm]
							                  	     
\S{5p'}{3/2}{2} & \S{5s}{3/2}{2}  & 556.22 & 0.11(1)  &   0.23  &   0.36  &   0.18  &        &   0.27  &   0.49  &  0.595  &  0.201     \\
                & \S{5s}{3/2}{1}  & 587.09 & 0.71(14) &   1.48  &   1.8   &   0.86  &        &   1.6   &   1.92  &  2.39   &  0.933     \\
                & \S{5s'}{1/2}{1} & 826.32 & 34.16(19)&   29.5  &   32.8  &         &   42.8 &   35.3  &   32.31 &  41.1   &  32.2      \\[2mm]
							                  	     
\S{5p'}{1/2}{1} & \S{5s}{3/2}{2}  & 557.03 & 0.980(56)&   1.71  &   2.4   &   1.37  &        &   1.9   &   3.18  &  3.91   &  1.32      \\
                & \S{5s}{3/2}{1}  & 587.99 & 0.055(6) &         &   0.13  &   0.049 &        &   0.093 &   0.14  &  0.198  &  0.0848    \\
                & \S{5s'}{1/2}{0} & 785.48 & 20.41(50)&   19.5  &   20.6  &         &        &   21.2  &   20.27 &  25.4   &  18.0      \\
                & \S{5s'}{1/2}{1} & 828.11 & 14.18(50)&   15.6  &   13.3  &         &   19.5 &   14.2  &   14.62 &  18.8   &  14.7      \\[2mm]
							                  	     
\S{5p'}{3/2}{1} & \S{5s}{3/2}{2}  & 567.25 & 0.015(3) &         &   0.02  &   0.022 &        &   0.02  &   0.01  &  0.00455&  0.00168   \\
                & \S{5s}{3/2}{1}  & 599.39 & 0.050(7) &         &   0.09  &   0.052 &        &   0.07  &   0.06  &  0.0841 &  0.0397    \\
                & \S{5s'}{1/2}{0} & 805.95 & 15.83(49)&   15.8  &   15.7  &         &   22.6 &   17.5  &   16.44 &  21.2   &  15.8      \\
                & \S{5s'}{1/2}{1} & 850.89 & 18.11(51)&   20.0  &   18.3  &         &        &   19.1  &   16.37 &  20.6   &  17.1      \\[2mm]
							                  	     
\S{5p}{1/2}{0}  & \S{5s}{3/2}{1}  & 758.74 & 43.1(6)  &   42.8  &         &         &   58.2 &   43.8  &   42.8  &  52.4   &  33.8      \\
                & \S{5s'}{1/2}{1} & 1212.35& 0.150(15)&         &         &         &        &   0.04  &   0.01  &  0.00143&  0.000407  \\[2mm]
							                  	     
\S{5p}{3/2}{2}  & \S{5s}{3/2}{2}  & 760.15 & 27.32(18)&   25.8  &   27.3  &         &   35.6 &   28.6  &   30.28 &  38.6   &  25.5      \\
                & \S{5s}{3/2}{1}  & 819.01 & 8.94(22) &   9.15  &   9.3   &         &   12.5 &   10.4  &   9.23  &  11.6   &  8.88      \\
                & \S{5s'}{1/2}{1} & 1373.89& 0.31(1)  &         &         &         &        &   0.3   &   0.13  &  0.164  &  0.356     \\[2mm]
							                  	     
\S{5p}{3/2}{1}  & \S{5s}{3/2}{2}  & 769.45 & 4.27(11) &   4.7   &   4.4   &         &        &   4.41  &   5.01  &  6.52   &  4.42      \\
                & \S{5s}{3/2}{1}  & 829.81 & 29.31(18)&   26.3  &   29.5  &         &        &   29.0  &   28.40 &  36.0   &  28.3      \\
                & \S{5s'}{1/2}{0} & 1286.19& 0.076(5) &         &         &         &        &   0.10  &   0.03  &  0.0345 &  0.0664    \\
                & \S{5s]}{1/2}{1} & 1404.57& 0.106(6) &         &         &         &        &   0.13  &   0.05  &  0.0633 &  0.141     \\[2mm]
							                  	     
\S{5p}{5/2}{2}  & \S{5s}{3/2}{2}  & 810.44 & 8.96(29) &   11.0  &   10.9  &         &   11.7 &   8.16  &   9.99  &  12.6   &  9.5       \\
                & \S{5s}{3/2}{1}  & 877.68 & 22.17(29)&   22.8  &   20.2  &         &        &   22.9  &   19.59 &  26.1   &  22.2      \\
                & \S{5s'}{1/2}{1} & 1547.40& 0.081(4) &         &         &         &        &   0.09  &   0.04  &  0.0477 &  0.131     \\[2mm]
							                  	     
\S{5p}{3/2}{3}  & \S{5s}{3/2}{2}  & 811.29 & 36.10(9) &   28.9  &         &         &   39.0 &   34.8  &   35.03 &  44.6   &  33.7      \\[2mm]
							                  	     
\S{5p}{1/2}{1}  & \S{5s}{3/2}{2}  & 892.87 & 22.89(65)&   30.7  &   26.2  &         &        &   19.8  &   22.30 &  28.4   &  26.1      \\
                & \S{5s}{3/2}{1}  & 975.18 & 3.13(14) &         &   3.4   &         &        &   4.28  &   2.63  &  3.43   &  3.75      \\
                & \S{5s'}{1/2}{0} & 1672.65& 0.126(6) &         &         &         &        &   0.15  &   0.05  &  0.0617 &  0.201     \\
                & \S{5s'}{1/2}{1} & 1878.55& 0.074(3) &         &         &         &        &   0.17  &   0.03  &  0.0344 &  0.143     \\
\hline
\hline
\end{tabular}
\end{center}
$^{\rm 1}$From Table \ref{table_aik}\\
$^{\rm 2}$Length form\\
$^{\rm 3}$Velocity form
\end{table}

%
%
\newpage

\newpage

\newpage

\newpage

\newpage

\newpage

\end{document}